\documentclass[12pt,epsfig]{article}

\usepackage{cite}
\usepackage{epsfig}
\usepackage{graphics}
\usepackage{inputenc}
\inputencoding{latin1}

\def\pythia{{\sc Pythia}}
\def\herwig{{\sc Herwig}}

\def\C2q{C_2(Q)}
\def\tC2q{$\C2q$}
\def\f2q{f_2(Q)}
\def\tf2q{$\f2q$}
\def\as{\alpha_s}
\def\tas{$\as$}

\newcommand{\PLB}[3]{Phys.~Lett.\ {\bf B#1} ({#3}) {#2}}

\newcommand{\PRD}[3]{Phys.~Rev.\ {\bf D#1} ({#3}) {#2}}
\newcommand{\PRL}[3]{Phys.~Rev.\ Lett.\ {\bf #1} ({#3}) {#2}}

\newcommand{\Comp}[3]{Comput.~Phys.~Comm.\ {\bf #1} ({#3}) {#2}}

\newcommand{\e}{\mathrm{e}}

\newcommand{\alps}{\alpha_s}
\newcommand{\lam}{\Lambda_{\mathrm{QCD}}}
\newcommand{\deta}{\Delta \eta}
\newcommand{\SF}{{\cal S}}

\begin{document}

\sloppy

\begin{titlepage}

  \begin{flushright}
    LU TP 99--13\\
    CERN--TH/99--240\\
    MAN/HEP/99/3 \\
    MC-TH-99/09 \\
    August 1999
  \end{flushright}
  \begin{center}
    
    \vskip 10mm {\Large\bf\boldmath Hard Colour Singlet Exchange at the 
Tevatron} \vskip 15mm

    {\large Brian Cox}\\
    $^*$Dept.~of Physics and Astronomy, University of Manchester\\
    Manchester M13 9PL, England\\
    coxb@mail.desy.de
    \vskip 10mm

    {\large Jeff Forshaw}\footnote{On leave of absence from $^*$}\\ 
    Theory Division, CERN \\
    1211 Geneva 23, Switzerland\\
    forshaw@mail.cern.ch
    \vskip 10mm

    {\large Leif Lönnblad}\\
    Dept.~of Theoretical Physics 2, S\"olvegatan 14A\\
    S-223 62  Lund, Sweden\\
    leif@thep.lu.se

  \end{center}
  \vskip 0mm
\begin{abstract}

We have performed a detailed phenomenological investigation of the 
hard colour singlet exchange process which is observed at the Tevatron in
events which have a large rapidity gap between outgoing jets. 
We include the effects of multiple interactions to obtain a prediction for 
the gap survival factor. Comparing the data on the fraction of gap 
events with the prediction from BFKL pomeron exchange we find agreement
provided that a constant value of $\alps$ is used in the BFKL calculation. 
Moreover, the value of $\alps$ is in line with that extracted from 
measurements made at HERA. 

\end{abstract}

\end{titlepage}

\section{Introduction}
\label{sec:intro}
Diffractive scattering has yet to be understood within the framework of QCD.
The problem arises because diffractive processes usually depend upon physics 
at distances which are not small on the scale typical of strong interactions.

Diffractive events are often selected by looking for regions in 
rapidity which are void of activity. Rapidity gaps are correlated with 
diffractive scattering since diffraction takes place in the limit where the
invariant centre-of-mass energy is much larger than any other invariant.
The challenge is to understand the gap producing mechanism using QCD.

\begin{figure}[h] 
\centerline{\epsfig{file=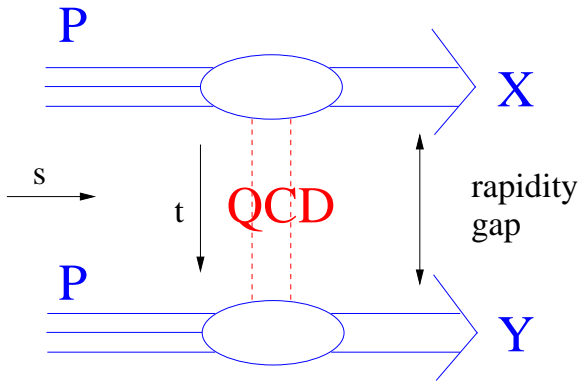,height=5.0cm,angle=0}}
\caption{A generic rapidity gap process.}
\label{rapdiagram}
\end{figure}

Figure \ref{rapdiagram} shows a generic rapidity gap event and serves
to set our notation.  The final state is composed of two clusters, $X$
and $Y$, which are separated in rapidity. As usual the Mandelstam
variables are labelled $s$ and $t$.

Typically, the gaps are produced with little or no momentum transfer, $t$, 
and this makes life difficult for those wanting to employ the 
tools of perturbative QCD. However, it is possible to squeeze the gap producing
mechanism to short distances: one looks for gap events in which a large
momentum is transferred across the gap, i.e. $-t \gg \lam^2$ \cite{Bj,MT}.

Experiments at the Tevatron and at HERA have already examined a variety of
these large $t$ diffractive processes, the most inclusive of which is
the double dissociation process $\gamma p \to X~Y$ which has been studied at 
HERA \cite{H1:dd}.\footnote{At HERA system $X$ is the product of the photon
dissociation and system $Y$ is the product of the proton dissociation.} 
The gaps-between-jets process, in which 
systems $X$ and $Y$ are each required to contain at least one jet, have been 
studied both at the Tevatron \cite{D0,D01,CDF,CDF1} and at HERA 
\cite{ZEUS,H1}. Additionally, the vector meson production process, 
$\gamma p \to V~Y$, in which the 
system $X$ is a single vector meson, $V$, has been measured at HERA 
\cite{H1:VM,ZEUS:VM}.

Theoretically, the presence of a large momentum transfer across the gap
supports the use of perturbative QCD \cite{FS}. The non-perturbative
physics can be factorised into the usual parton density functions which
can then be convoluted with the hard partonic scattering subprocess to
determine the physical cross-section. The hard scattering subprocess
can be calculated to an accuracy where all leading logarithms in
energy, i.e. all terms $\sim \alps^n \ln^n (s/|t|)$, are summed up using the 
formalism of BFKL \cite{BFKL}. 
Calculations have been performed for all of the above
processes to leading logarithmic accuracy, see \cite{JF:review} for 
a recent summary. 

In this paper, we wish to examine the Tevatron data on the gaps-between-jets 
process. We perform a detailed analysis of these data at 630~GeV and 1800~GeV
and are drawn to conclude that the data are fully consistent 
with the leading logarithmic
calculation of BFKL providing one fixes the strong coupling.
This conclusion is consistent with that which can also be drawn 
from the HERA data collected to date. In our study, we pay particular 
attention to the issue of gap survival: we use \pythia\ to model the 
underlying event \cite{pythia}. We consider observables at both the 
parton and hadron level. 

\subsection*{The theoretical calculation}
We start by explaining how we perform our theoretical calculations.
We use the hard partonic subprocess first calculated in \cite{MT}:
\begin{equation}
\frac{d \sigma(q q \to q q)}{dt} = \left( \frac{\alps C_F}{\pi} \right)^4
\frac{16 \pi^3}{(N_c^2-1)^2} \frac{1}{t^2} \left[ \int d\nu
\frac{\nu^2}{(\nu^2+1/4)^2} \e^{\omega(\nu) y} \right]^2 \label{exact}
\end{equation}
where the leading logarithm BFKL kernel is
\begin{equation}
\omega(\nu) = 2 \frac{C_A \alps}{\pi} {\rm Re}[ \psi(1) - \psi(1/2+i \nu) ]
\label{kernel}
\end{equation}
and
\begin{equation}
y = \deta = \ln \left( \frac{\hat{s}}{-t} \right).
\end{equation}
To this level of accuracy, the jets have equal and opposite transverse 
momenta, $p_T^2 = -t$. Throughout this paper we evaluate the right-hand-side 
of (\ref{exact}) without approximation although we do note
that, in the limit $\deta \gg 1$, it simplifies to
\begin{equation}
\frac{d \sigma(q q \to q q)}{dt} \approx (C_F \alps)^4 \frac{2\pi^3}{t^2}
\frac{\exp(2 \omega_0 y)}{(7 \alps C_A \zeta(3) y)^3}
\label{BFKL:asym}
\end{equation} 
where $\omega_0 = \omega(0) = C_A (4 \ln 2/\pi) \alps$.
The gluon-quark and gluon-gluon subprocesses are the same as the quark-quark
subprocess, up to colour factors, and so we can write the parton level 
cross-section for the gaps-between-jets process as
\begin{eqnarray}
\frac{d \sigma(h_1~h_2 \to X~Y)}{dx_1 dx_2 dt} &=& 
\left( \frac{81}{16}g_1(x_1,\mu^2) + \Sigma_1(x_1,\mu^2) \right)
\left( \frac{81}{16}g_2(x_2,\mu^2) + \Sigma_2(x_2,\mu^2) \right)
\nonumber \\ & &
\frac{d \sigma(q q \to q q)}{dt} \label{lla}
\end{eqnarray}
where $g_i(x_i,\mu^2)$ is the gluon parton density function for hadron $i$
and $\Sigma_i(x_i,\mu^2)$ is the sum over all quark and antiquark density
functions for hadron $i$. These sub-processes have been coded into \herwig\ 
\cite{herwig,JMB} thereby allowing one to compute the likely corrections due
to parton showering and hadronisation. We use a particular version of 
\herwig\ which includes the BFKL subprocesses obtained using (\ref{exact})
without approximation \cite{Hayes}.\footnote{The code can be obtained on
request from \texttt{coxb@mail.desy.de}.}
Subsequently we take the factorisation 
scale $\mu^2 = -t$ (in the region of interest, $s \gg -t$, this is a very good 
approximation to the value chosen in \herwig). We have used
both CTEQ2M and CTEQ3M parton distribution functions \cite{CTEQ,PDFLIB}, and 
have found the differences to be very small.

\begin{figure}[h] 
\centerline{\epsfig{file=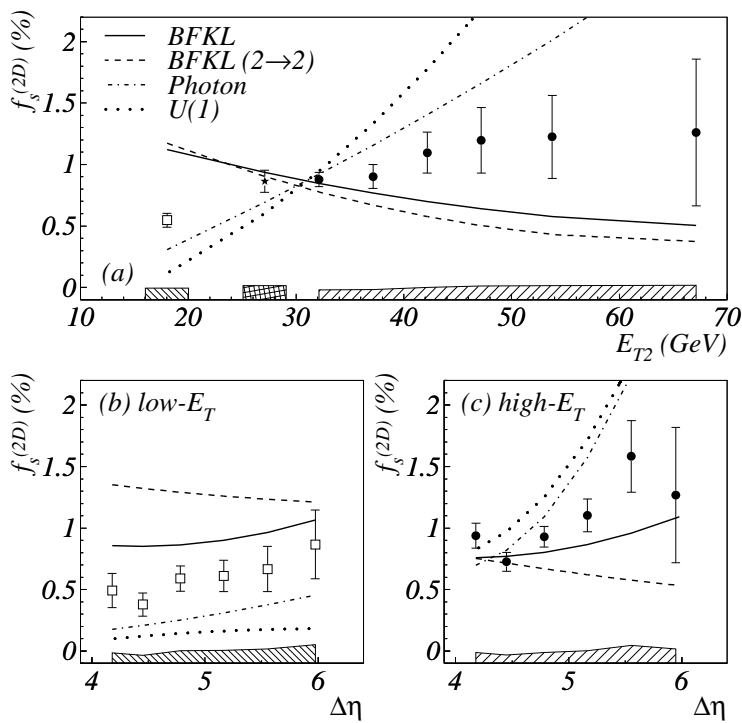,height=9.5cm}}
\caption{D\O\ data compared with a BFKL calculation. Plot from \cite{D01}.}
\label{D0:data}
\end{figure}

\subsection*{D\O\ data versus the BFKL pomeron}
Let us now summarise the situation to date. Figure \ref{D0:data}
shows the D\O\ data compared to a BFKL calculation. What is shown is the
fraction of all dijet events in which the central region of the detector 
$|\eta| < 1$ is devoid of hadronic activity (see later for a more detailed 
explanation of the D\O\ selection cuts) as a function of the transverse energy 
of the lower $E_T$ jet, $E_{T2}$, and the separation in rapidity of the
jets, $\deta$. The CDF data at 1800 GeV do not go down to 
such low values of $E_T$ and are consistent with a flat $E_{T}$ distribution.

The BFKL curve is quite clearly excluded by the data. However, it should 
be understood that this curve was generated using the default settings in
\herwig \cite{JMB}. 
In particular, this means that the asymptotic cross-section of
(\ref{BFKL:asym}) was used with a fixed value of $\omega_0 = 0.3$ in the 
exponent and a fixed value of $\alps = 0.25$ in the denominator, whilst the 
$\alps^4$ prefactor was allowed to run with $-t$ according to the two-loop
beta function. These settings are specific to the BFKL subprocess, in all
other subprocesses the \herwig\ default is to use the two-loop running
coupling. The falling of the BFKL curve with increasing $E_{T2}$ is 
driven by the running of the coupling in the pre-factor since the gap 
fraction goes like $\sim \alps^4 / \alps^2$. 

Now, to leading logarithmic accuracy $\alps$ is simply an unknown parameter.
Higher order corrections will indeed cause the coupling to run, however it
is not clear how this should be done in a consistent way. In this paper
we restrict ourselves to the leading logarithmic approximation and treat the
coupling as a free parameter. Moreover, we are guided by recent HERA data
on the double dissociation process \cite{H1:dd} which can be described by the
leading logarithmic BFKL formalism with a fixed strong coupling, 
$\alps = 0.17$. We also note that a fixed coupling constant was needed in
order to explain the high-$t$ data on $p\bar{p}$ elastic scattering 
via three gluon exchange \cite{DL}. 
Furthermore, NLO corrections suggest a fixed value for 
the leading eigenvalue of the BFKL equation, $\omega(0)$, \cite{Brodsky} 
which in turn suggests the use of a fixed coupling in the LLA kernel, i.e. in 
equation (\ref{kernel}).
    
\subsection*{Underlying events and gap survival}

So far we have said nothing about the possibility that gaps produced by the
colour singlet exchange might be filled in as a consequence of secondary
interactions between the colliding hadrons. The possibility that gaps do not
survive formally invalidates the factorisation of the cross-section which we
assumed in the previous sub-section. However, there are fairly good reasons
(as we shall see in the next section) for expecting the effect to enter as a 
multiplicative `gap survival' factor, $\SF$, which specifies the probability 
that a gap produced at the hard scatter level survives to the hadron level. 
Note that we do expect $\SF$ to vary with 
the centre-of-mass energy of the colliding beams: as the centre-of-mass energy
increases the corresponding proliferation of low-$x$ partons makes a 
secondary partonic interaction more likely. 

In the next section we shall use \pythia\ to extract the gap survival
factors relevant to the Tevatron operating at 1800~GeV, at 630~GeV and for
HERA at a fixed $\gamma p$ energy of~200 GeV. We will then use these 
factors to correct our cross-sections for processes involving gaps, i.e.
\begin{equation}
\frac{d \sigma(h_1~h_2 \to X~Y)}{dx_1 dx_2 dt} \to 
\SF \frac{d \sigma(h_1~h_2 \to X~Y)}{dx_1 dx_2 dt},
\end{equation} 
where $\SF$ is the fraction of gap events which survive after including the
effect of the underlying event as modelled in \pythia.

\section{Gap Survival and Multiple Interactions}
\label{sec:mi}

To investigate the gap survival probabilities we use a model for
multiple interactions available in the \pythia\ program. Here the
probability to have several parton--parton interactions in the same
collision is modelled using perturbative QCD. For a given hard
parton--parton scattering there is a probability of having additional
scatterings given by the parton densities and the usual LO
$2\rightarrow 2$ matrix elements. The matrix elements are divergent
for $p_\perp\rightarrow 0$ and are regulated by replacing the
$1/p_\perp^4$ pole with $1/(p_{\perp}^2+p_{\perp 0}^2)^2$ and using
$p_{\perp}^2+p_{\perp 0}^2$ as argument in \tas\ rather than just
$p_\perp^2$. The value of $p_{\perp 0}$ varies with collision energy,
$E_{{\rm{cm}}}$, and we have used the default behaviour: $p_{\perp 0} =
( E_{{\rm{cm}}}/(1\mbox{TeV}) )^{0.16} \cdot 2.1$~GeV.

\begin{figure}[ht]
\begin{minipage}[t]{0.475\textwidth}
\centerline{\resizebox{8cm}{!}{\rotatebox{0}{\includegraphics{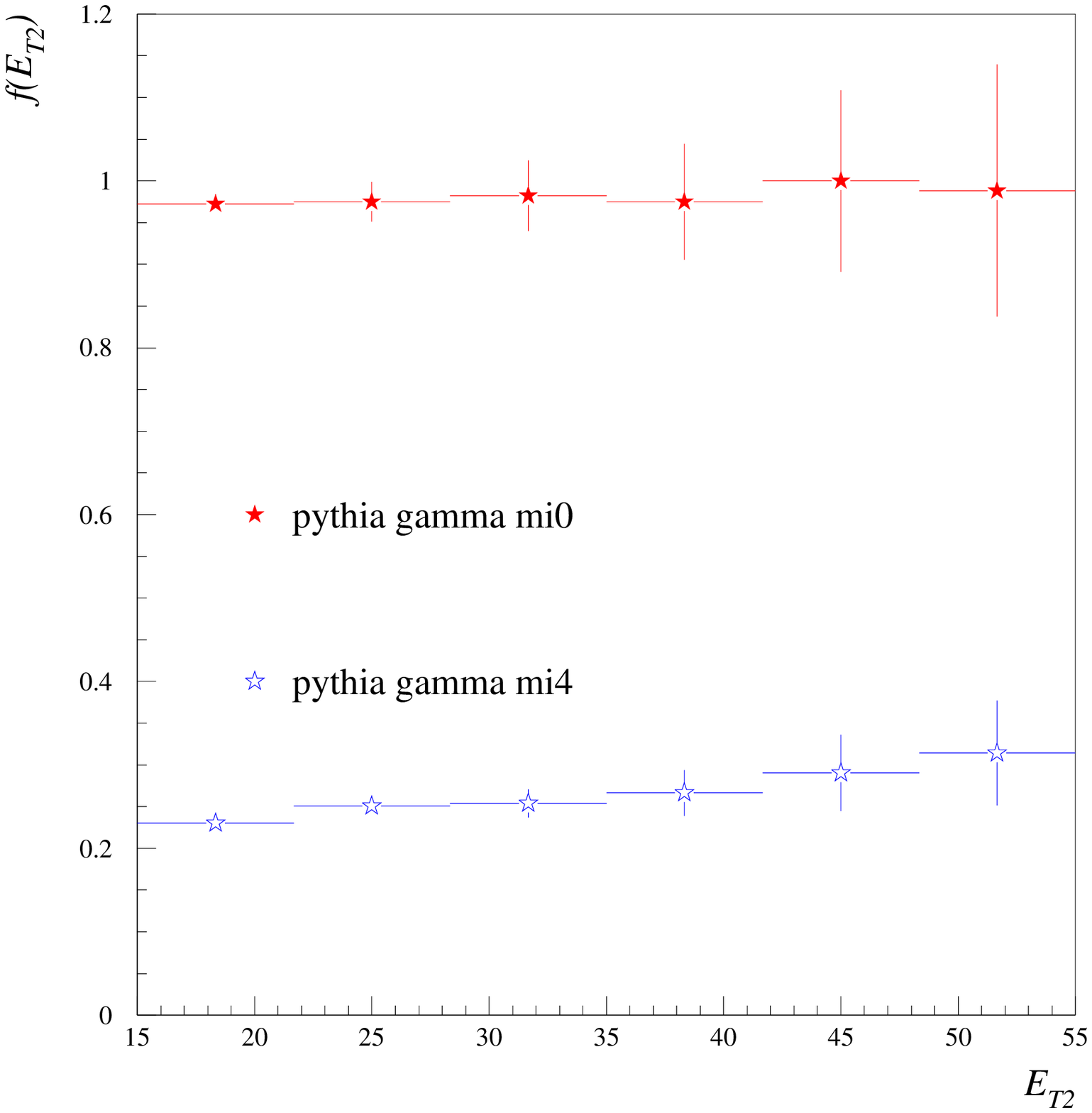}}}}
\caption{Gap fraction in photon exchange events at 1800~GeV with/without MI's:
$E_{T2}$ spectrum}
\label{gapfrac1}
\end{minipage}\hspace*{\fill}
\begin{minipage}[t]{0.475\textwidth}
\centerline{\resizebox{8cm}{!}{\rotatebox{0}{\includegraphics{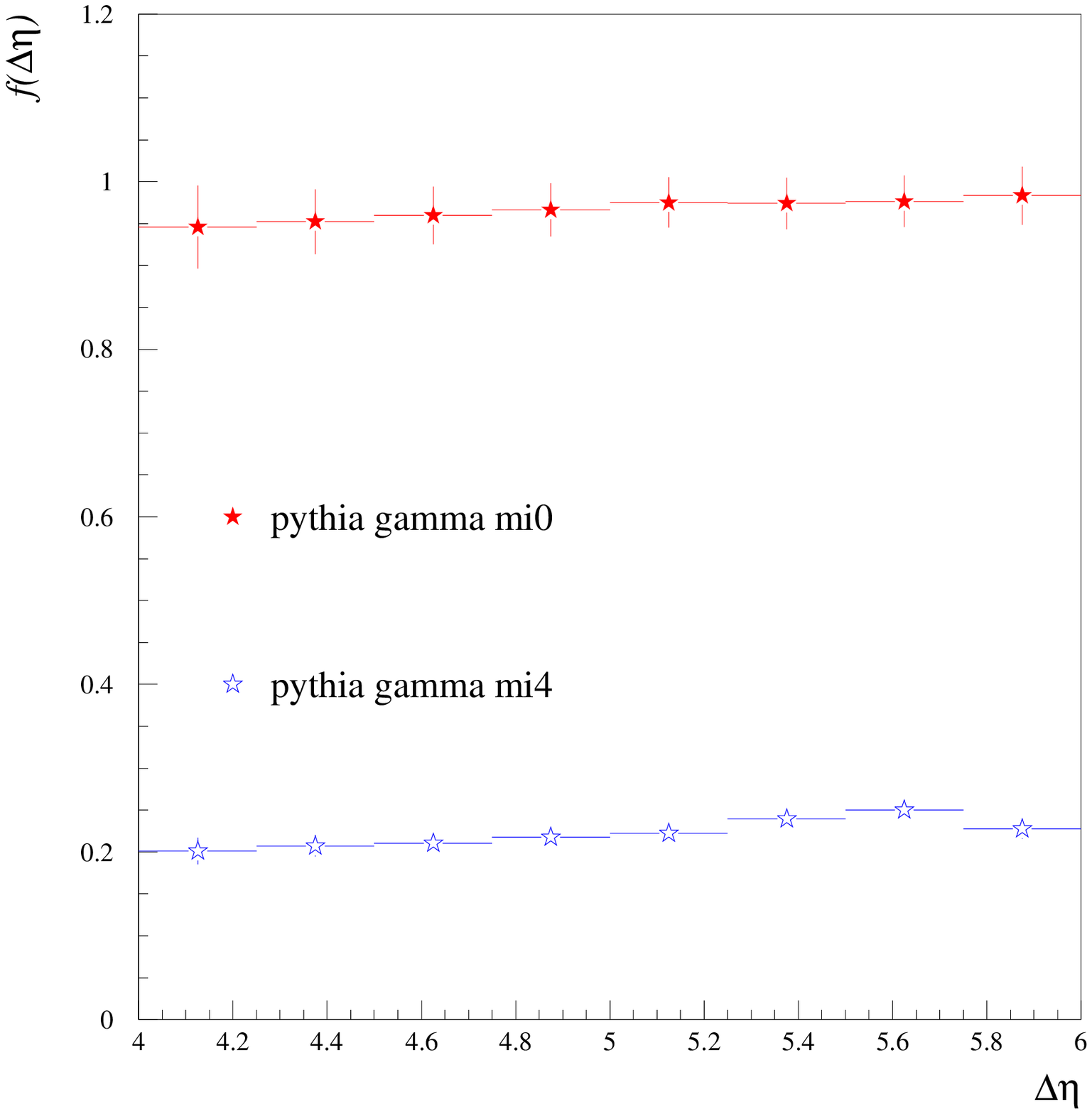}}}}
\caption{Gap fraction in photon exchange events at 1800 GeV with/without MI's:
$\deta$ spectrum}
\label{gapfrac2}
\end{minipage}
\end{figure}



\begin{figure}[ht]
\begin{minipage}[t]{0.475\textwidth}
\centerline{\resizebox{8cm}{!}{\rotatebox{0}{\includegraphics{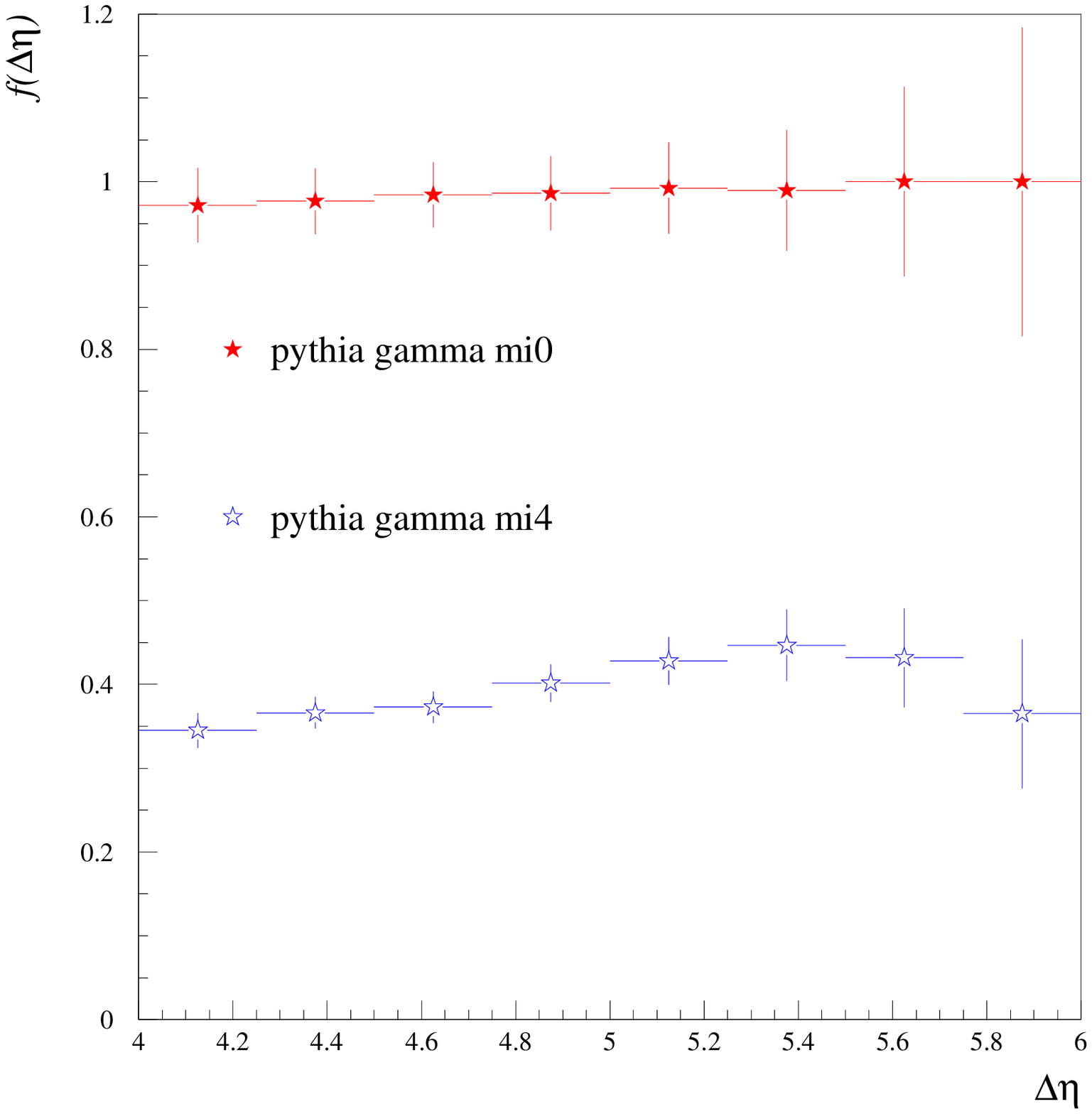}}}}
\caption{Gap fraction in photon exchange events at 630 GeV with/without MI's}
\label{gapfrac3}
\end{minipage}\hspace*{\fill}
\begin{minipage}[t]{0.475\textwidth}
\centerline{\resizebox{8cm}{!}{\rotatebox{0}
{\includegraphics{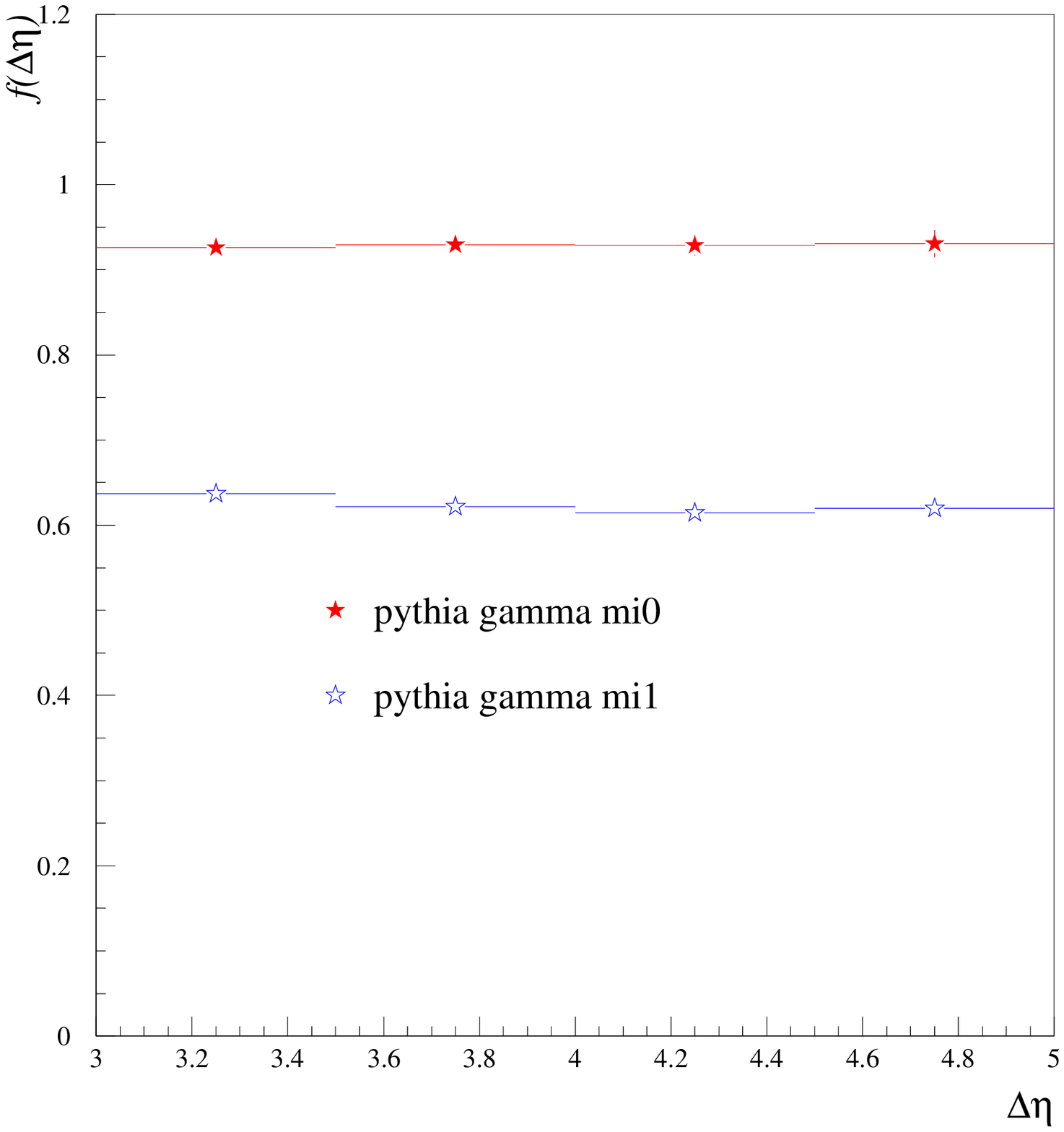}}}}
\caption{Gap fraction in photon exchange events at HERA with/without MI's}
\label{gapHERA}
\end{minipage}
\end{figure}



The probability for additional interactions is not fixed but varies
according to an impact-parameter picture, where central
collisions are more likely to have multiple interactions. The partons
in the proton are assumed to be distributed according to a double-gaussian as 
described in \cite{sjo87a,pythia}. There are several parameters in this model 
and we have used the default setting for each.\footnote{Setting
  the switch \texttt{MSTP(82)=4} in \pythia, with everything else
  default, will give the model as we have used it.}
This is adequate for the accuracy required here but it should be noted
that the parameters have not been tuned to Tevatron data.

Hard BFKL pomeron exchange has not been implemented in \pythia\ yet,
but we can investigate the effects of multiple interactions on the gap
survival probabilities in general by looking at high-$t$ photon
exchange. In Figure \ref{gapfrac1} and Figure \ref{gapfrac2} we show the gap 
fraction in photon exchange events at 1800~GeV with (mi4) and without (mi0) 
multiple interactions and, as expected, 
the probability for a gap introduced by the colourless photon
scattering is greatly reduced in the presence of multiple interactions.
The reason is that any additional scattering is likely to be a colourful one, 
destroying the gap. Thus we find the gap survival probability to 
be a simple multiplicative factor almost independent of the  
gap size and $E_T$ although varying with the total collision energy.
Figure \ref{gapfrac3} shows the effect of multiple interactions at the lower
energy of 630~GeV. 

In Figure \ref{gapHERA} we show the effect in $\gamma p$ 
interactions at 200~GeV, as is relevant for HERA. In \pythia\ the procedure 
for generating these $\gamma p$ events requires some care. The cross-section 
is divided into three parts: direct, anomalous and VMD. Direct interactions, 
where the photon couples directly to the hard interaction, are not present in
gamma exchange. Anomalous processes are resolved ones where the
evolution from the photon is purely perturbative, while VMD are
resolved processes where the photon has fluctuated into a vector meson
state. Multiple interactions are only included in the latter
processes, which is why the gap survival probability is much higher
than in the hadron-hadron case. For $\gamma p$ we also use a simpler approach 
to multiple interactions, with an impact-parameter-independent probability 
for secondary scatterings since for this option there exists a set of tuned 
parameters \cite{jon} which we have used in the generation. The gap fraction
after including multiple interactions is labelled `mi1' in 
Figure \ref{gapHERA}. 

In summary, we find $\SF(1800~{\rm GeV}) = 22 \%$, 
$\SF(630~{\rm GeV}) = 35 \%$ and $\SF({\rm HERA})~=~67\%$.

Note that requiring a large rapidity gap not only selects events with 
colour-singlet exchange in general, but in particular events with 
colour-singlet exchange without additional interactions.

\begin{figure}[h]
\centerline{\epsfig{file=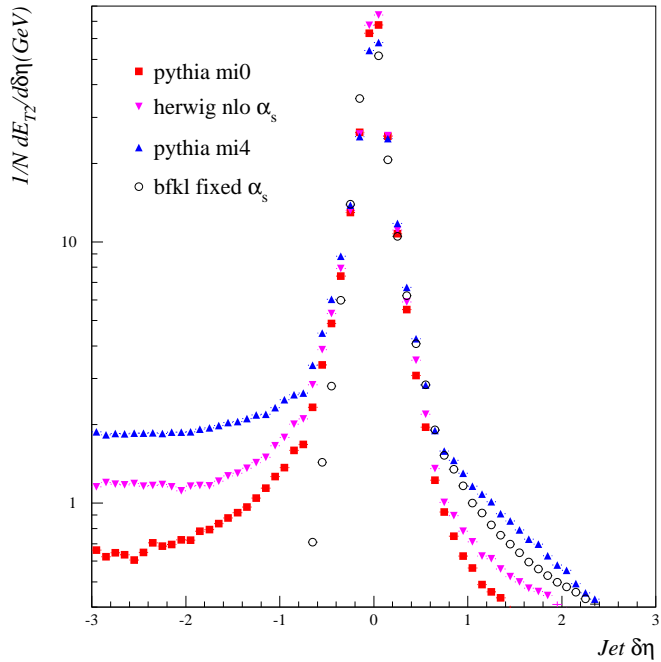,height=10cm}}
\caption{Jet $\eta$ profiles}
\label{profileseta}
\end{figure}

Multiple interactions also give rise to the so-called jet pedestal and
underlying event effects. This means that the jets measured in
hadron-hadron collisions cannot be compared directly to e.g.\ 
predictions from fixed order perturbation theory. In Figure
\ref{profileseta} we show jet profiles obtained from \pythia\ with and
without multiple interactions (and with $|\delta \phi| < 0.7$). 
The proton remnant is at $\delta \eta > 0$.
It is clear that multiple interactions introduce a jet pedestal of more 
than 1~GeV of $E_T$ per unit rapidity. For comparison, also shown is the jet 
pedestal from \herwig. We note that \herwig\ predicts a greater amount of 
energy outside the jet cone than \pythia\ without multiple interactions.
On the remnant side, $\delta \eta > 0$, the difference is accounted for by
different treatments of the remnant (running \herwig\ with \verb+IOPREM=0+
reproduces the \pythia\ result). However, on the gap side, $\delta \eta < 0$,
the large difference is not so easy to explain. \herwig\ does use a 
larger coupling than \pythia\ (see later), although this is not enough to 
account for the differences. The disagreement
could be related to the differences in treatment of QCD coherence. Coherence 
in \herwig\ is achieved by ordering emissions in the parton cascades in
angle, while in \pythia\ the ordering is in virtuality with an
additional angular constraint. The double ordering in \pythia\ reduces
the emission probability and the fact that we are looking at jets at large 
rapidity (small angles) may enhance this effect. We have checked that relaxing
the angular constraint in \pythia\ increases the $E_T$-flow on the gap
side by $\approx20\%$, although this needs to be studied further
before firm conclusions can be made. Fortunately, as we shall see, it
is likely that these differences mostly influence the overall normalisation 
of the gap fraction. However, it is important that these jet shapes are 
thoroughly confronted with data before precise estimates of the hadronisation
corrections can be made. 

In the D\O\ jet measurements the excess $E_T$ from the underlying event
is taken into account by correcting the jet $E_T$ using data. 
In particular, the correction
is determined by looking at the $E_T$ in regions away from the jets.
The correctness of this subtraction is also supported by a study of
minimum bias events. The bottom line is that D\O\ subtract around 
1~GeV from the $E_T$ of each reconstructed jet \cite{D0jets}.
In particular, in the gap fraction measurement, this subtraction is
performed for all jets, including those in gap events. But, as we already 
pointed out, requiring a large rapidity gap also selects events without 
multiple interactions, where the jet pedestal is absent, or at least much
smaller. Now, 1~GeV may seem like a small correction when we are looking
at jets of $E_T=18$~GeV or larger, but recall that the $E_T$
spectrum is very steeply falling: typically it falls faster than
$1/E_T^4$. The correction may therefore be as large as 30\% for
$E_T=18$~GeV, while for $E_T>50$~GeV it will be less than
10\%. Looking at Figure \ref{D0:data} we can roughly undo the
erroneous correction of the jets in the gap events. The lowest
$E_T$ bin would then be increased a factor $(19/18)^4\approx 1.24$
while the other points would be less affected. The increase of the gap 
fraction with increasing $E_T$ then becomes less pronounced.

\section{Results}
\subsection*{Cross-sections at the parton and hadron level}
In this section we show our results for jet cross-sections. Specifically, 
we work at 1800~GeV centre-of-mass energy and make the same
analysis cuts as the D\O\ collaboration \cite{D01}. 
Jets are found using a cone algorithm 
\cite{pxcone,cone} with cone radius $0.7$ and the \verb=OVLIM= 
parameter set to $0.5$. 
The inclusive dijet sample is defined by the following cuts:
\begin{itemize}
\itemsep 0mm
\item{$|\eta_1|, |\eta_2| > 1.9$, i.e. jets are forward or backward}
\item{$\eta_1 \eta_2 < 0$, i.e. opposite side jets}
\item{$E_{T2} > 15$ GeV}
\item{$\deta > 4$, i.e. jets are far apart in rapidity}.
\end{itemize}
The sub-sample of gap events is obtained by employing the further cut that
there be no particles emitted in the central region $|\eta| < 1$ with
energy greater than 300 MeV. 
The CDF collaboration have performed a similar analysis \cite{CDF1} and we 
comment on their data wherever relevant. 
CDF use the same selection criteria as D\O\
except that they chose $1.8 < |\eta_{1,2}| < 3.5$, $E_{T2} > 20$ GeV for the 
1800 GeV sample and $E_{T2} > 8$ GeV for the 630 GeV sample, and they    
make no further cut on the jet separation.\footnote{Strictly speaking we
should set \texttt{OVLIM=0.75} when comparing to CDF data. However, the 
effect of implementing this change is negligible.}  

We note that observables like the fraction of gap events to all events (the
so-called gap fraction) are strictly not infrared safe. The definition of
a gap event requires the specification of some threshold energy. Particles
emitted into the gap with energy below this threshold by definition do not 
spoil the gap. In theoretical calculations this leads to logarithms in the 
infrared scale which formally diverge as the threshold is taken to zero. 
Thus theoretical predictions based on perturbation theory are in principle 
unstable as the threshold is lowered. One way to avoid this problem is to 
define a `gap' to be an interval in rapidity which does not contain any jets 
above some $E_T$ which is perturbatively large. However, the CDF and D\O\
gap samples are defined in such a way as to greatly reduce the sensitivity
to the threshold energy. By insisting that the gap be a strip of width at
least 2 units in rapidity centred on $\eta = 0$ and by forcing the jets to be 
at $\eta > 1.9$ the majority of events have gaps which do not encroach too
closely to the edge of the jets. As a result, any soft radiation tends not
to be emitted into the gap (it is localised around the leading parton). The
problem would have been more acute if the gap sample had been defined by
insisting that no particles be emitted (above the threshold) between the 
edges of the jet cones.
This latter protocol has been used by the HERA experiments and should lead
to results which depend significantly upon the threshold energy. We have 
explicitly checked that our results for the Tevatron do not change 
significantly as the threshold energy is varied from 0 MeV to 300 MeV.
  
In all cases results labelled `bfkl fixed $\alps$' represent the cross-sections
obtained using only the $2 \to 2$ hard subprocess cross-section of
(\ref{exact}) with $\alps = 0.17$; the value hinted at by the HERA data. 
Those labelled with `bfkl running $\alps$' are obtained 
using the running coupling with the two-loop beta function 
($\lam^{(5)} = 200$~MeV) in the $\alps^4$ prefactor whilst keeping a fixed 
coupling, $\alps = 0.17$, in the evaluation of $\omega(\nu)$. The parton
showering and hadronisation are determined by the usual \herwig\ defaults.
Note that we have not yet corrected the BFKL data by the gap survival factor
determined earlier.

\begin{figure}[h] 
\centerline{\epsfig{file=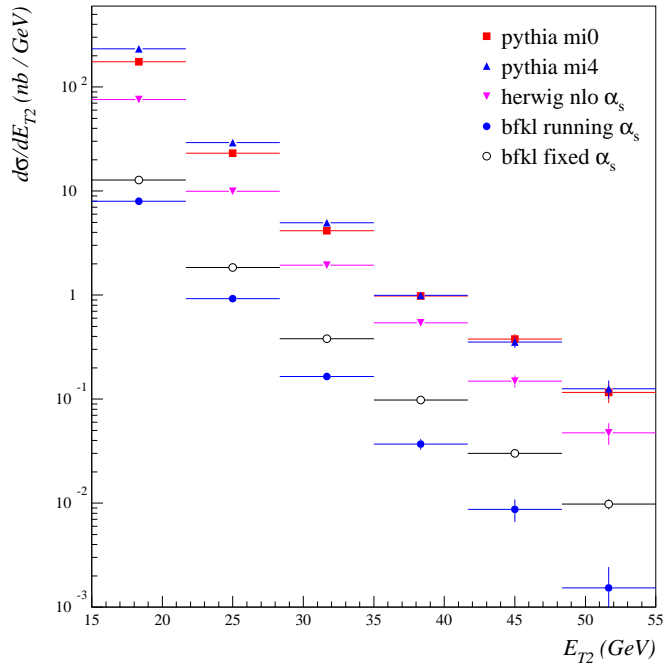,height=10cm}}
\caption{Cross-sections at the hadron level}
\label{et}
\end{figure}

All other results are cross-sections for the non-BFKL $2 \to 2$ 
subprocesses, i.e. they correspond to the usual QCD non-colour-singlet 
exchange subprocesses. More specifically: curves labelled 
`\herwig\ lo' are generated by \herwig\ with a one-loop running coupling 
($\lam^{(5)} = $ 362~MeV) whilst those labelled `\herwig\ nlo' are the 
\herwig\ default curves obtained using a two-loop running coupling 
($\lam^{(5)} \approx$ 200~MeV); similarly the \pythia\ curves are generated 
without multiple interactions (`mi0') or with multiple interactions (`mi4') 
and correspond to a one-loop running coupling with $\lam$ determined by the
value used in fitting the chosen parton density functions 
($\lam^{(5)} \approx$ 200~MeV when using CTEQ2M). 
The $\lam$ values quoted are the ones used to generate the parton level 
theory curves shown on some of the plots
and, at two loop, are quoted in the $\overline{MS}$ scheme. The
parton level theory curves are obtained from those bare $2 \to 2$ subprocess
cross-sections which involve $t$-channel gluon exchange ($t$-channel quark 
exchange gives a negligible contribution), i.e. they do not include the
effects of parton showering and hadronisation. 

\begin{figure}[h] 
\centerline{\epsfig{file=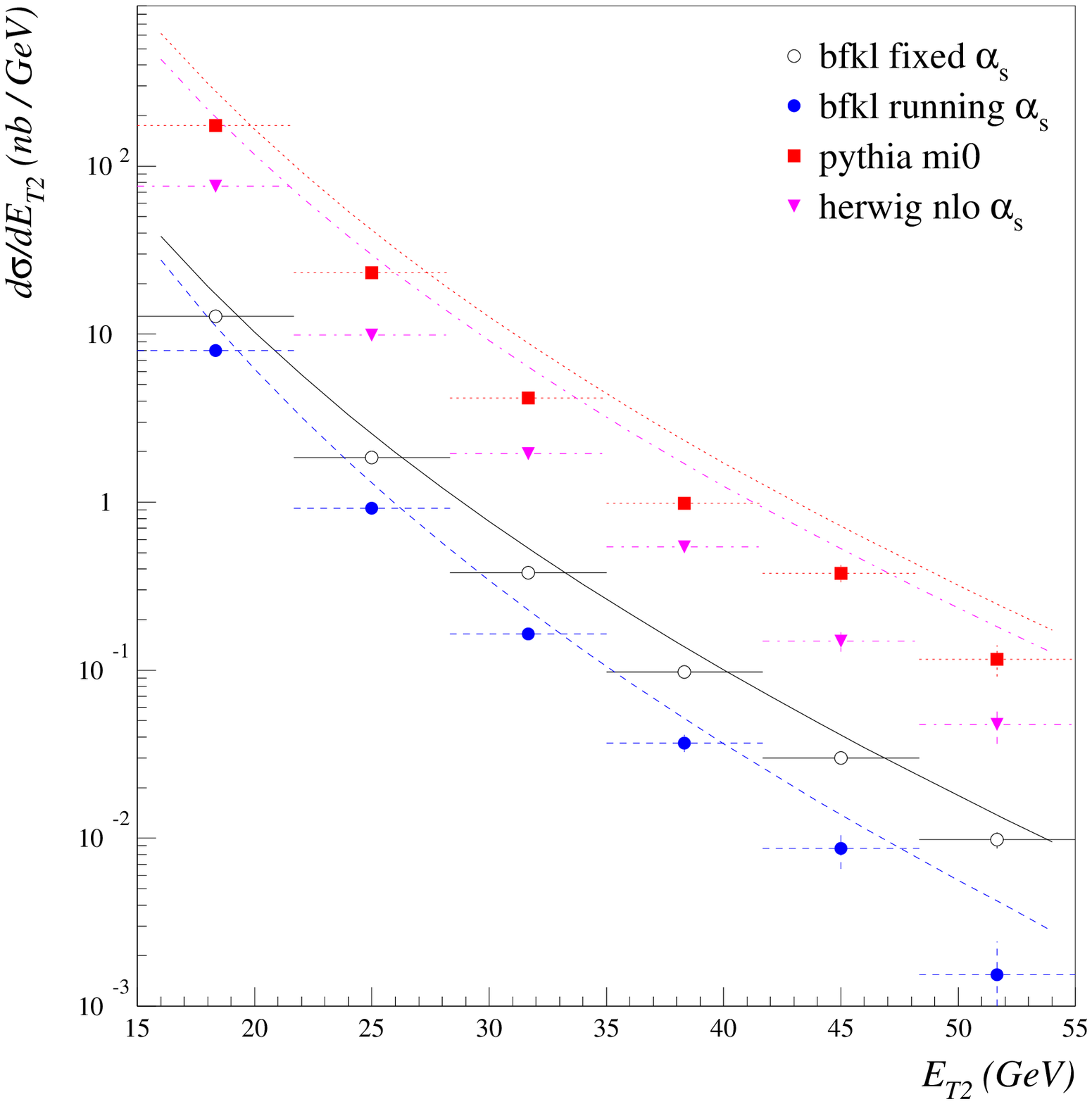,height=10cm}}
\caption{Comparison of Monte Carlo hadron level with theory parton level}
\label{etth}
\end{figure}

In Figure \ref{et} we show the $E_T$ spectrum of the lower $E_T$ jet after
hadronisation and parton showering. Note that there is a significant 
difference between the \herwig\ and \pythia\ predictions. 
This is due to the different ways of running the coupling employed in 
the two programmes and the differences in fragmentation discussed earlier: 
data would help resolve the issue.
Further sensitivity to the choice of $\alps$ (even in the 
non-BFKL case) can be seen in Figure \ref{etth}, where we show the 
\pythia\ $E_T$ spectrum of the 
lower $E_T$ jet after hadronisation and parton showering but without 
multiple interactions, and compare to the parton level theory predictions 
and to \herwig. Hadronisation and parton showering corrections are clearly 
very significant in both the BFKL and non-BFKL data sets. We have also
looked at the relative role of hadronisation and parton showering and find
that each contributes significantly to the deviations from the parton level
theory predictions. 

\begin{figure}[h] 
\centerline{\epsfig{file=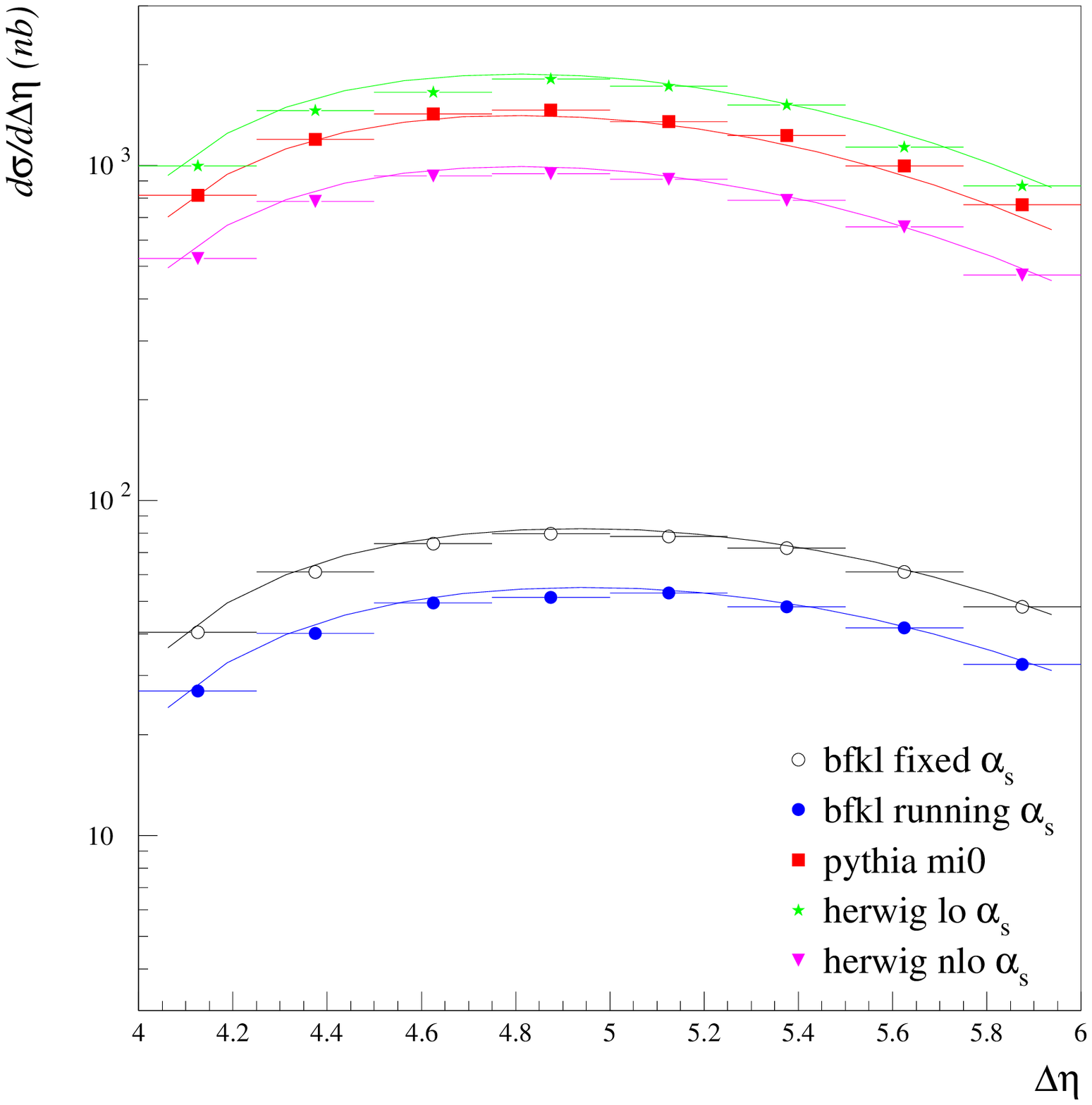,height=10cm}}
\caption{Low $E_T$ events. Comparison of Monte Carlo parton level with 
theory parton level}
\label{deta25}
\end{figure}

In Figure \ref{deta25} we compare the parton level theory curves to the
parton level cross-sections obtained from Monte Carlo. These $\deta$
distributions are obtained for the `low $E_T$' jet sample, i.e. for those
jets in the range 15~GeV $< E_{T2} <$ 25~GeV. 
As well as providing a cross-check, they serve to illustrate the dominance 
of $t$-channel gluon exchange. This plot also illustrates the effect of 
insisting that $|\eta_1|,~|\eta_2| > 1.9$. 
The fall of the curve at low $\deta$ is a result of the increasingly strong
requirement that the incoming partons have nearly equal and opposite energies.
As a result, in the absence of the $\deta>4$ cut, these cross-sections would
fall smoothly to zero at $\deta=3.8$. This kinematic effect could be removed
by relaxing the constraint that the gap be central, i.e. by allowing the 
whole dijet system to tilt. The fall off of the $\deta$ distribution at large 
$\deta$ is due to the decrease of the parton densities with increasing $x$.

\subsection*{Gap fractions}
We now turn our attention to the gap fraction. This is the quantity usually
measured by experiment. To construct the gap fraction we form the ratio
\begin{equation}
f = \frac{ N_{BFKL}^g + N_{QCD}^g } 
{ N_{BFKL} + N_{QCD} }
\end{equation}
where $N_{BFKL}^g$ is the number of events in the BFKL sample which
also satisfy the gap cut, $N_{QCD}^g$ is the number of events in
the non-BFKL sample which satisfy the gap cut and the sum in the denominator
is over all events in the dijet sample. The numerator is overwhelmingly
dominated by BFKL events in all the plots we show.

\begin{figure}[h] 
\centerline{\epsfig{file=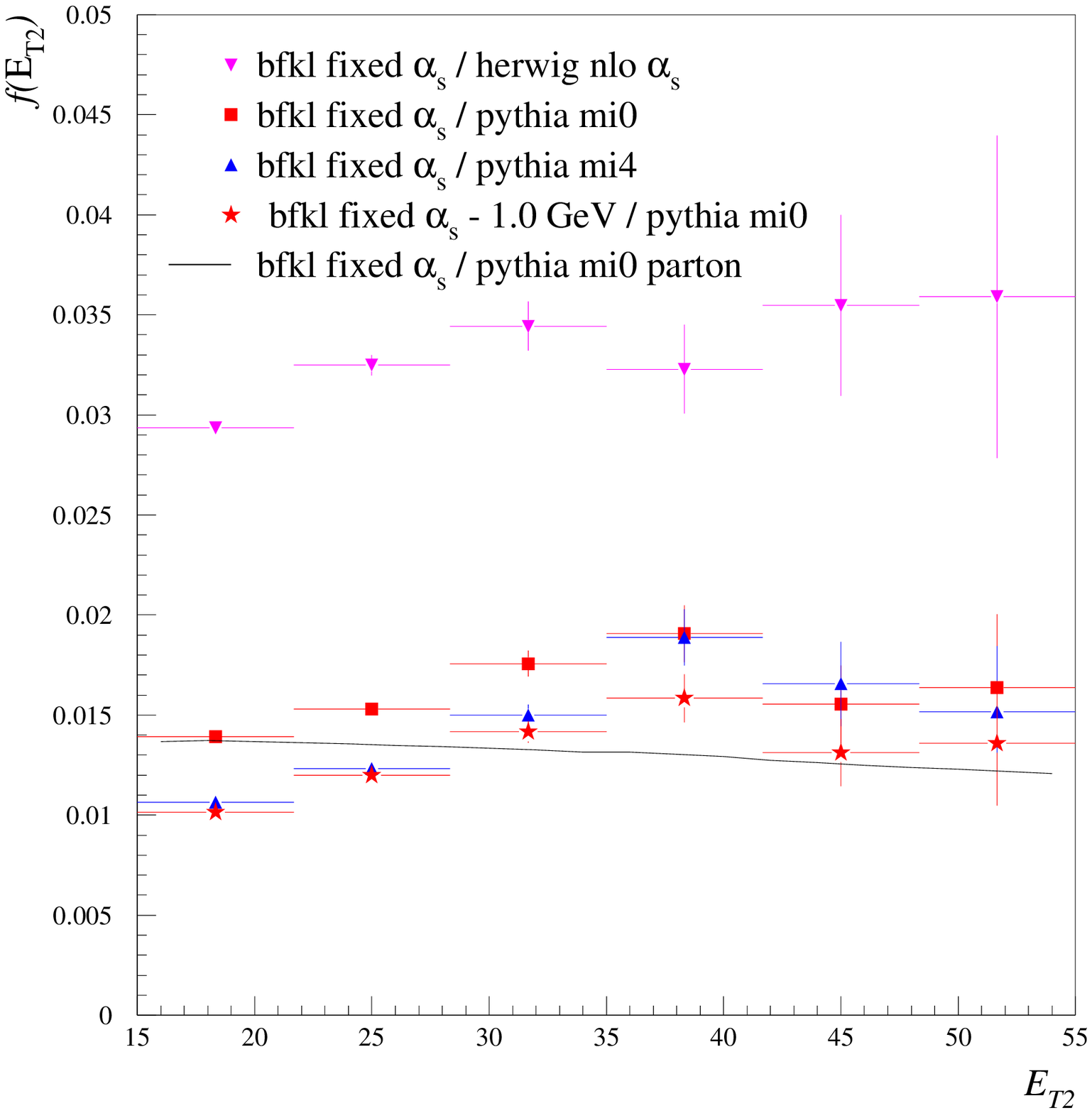,height=10cm}}
\caption{Gap fraction as a function of $E_{T2}$}
\label{gfet}
\end{figure}

In Figure \ref{gfet} we show the hadron level gap fraction as a function of 
$E_{T2}$. All our gap fraction plots include the \pythia\ gap survival 
correction factor. The stars, corresponding to the ratio 
`bfkl~fixed~$\alps~-1$~GeV~$/$~pythia~mi0', are the result of simulating 
the D\O\ jet
correction procedure. We have subtracted 1~GeV from the BFKL jets and 
generated the non-BFKL sample by running \pythia\ without multiple 
interactions. Notice that these data points are in agreement with the 
the triangles labelled `bfkl~fixed~$\alps/$~pythia~mi4' 
which have been obtained
without applying the jet correction to the BFKL sample and including multiple
interaction effects in the non-BFKL sample. This 
agreement confirms our earlier claim that if one should want to correct for
the affect of the underlying event (i.e. multiple interactions) then the
correction can be approximated by a subtraction of 1~GeV from each jet. 
The squares in Figure \ref{gfet} correspond to the gap fraction properly 
corrected for the underlying event, i.e. only the denominator (non-gap sample)
has been corrected. Also shown in Figure \ref{gfet} are comparisons with 
the ratio predicted using \herwig\ to generate the non-BFKL sample and with
a parton level theory calculation, labelled `pythia mi0 parton' which was 
computed using the same prescription as \pythia\ for the running coupling 
(as detailed above).

The rise at low $E_{T2}$ is mainly a result of including multiple interactions
since, at low $E_{T2}$, the small amount of energy arising from the 
underlying event produced by secondary scatters enhances significantly the 
denominator (i.e. the full dijet sample) thereby reducing the gap fraction.
Although we do note that a weaker rise, due to parton showering and
hadronisation corrections, is still present in the \herwig\ and
\pythia\ mi0 samples. 

Note also the decline of this rise of the gap fraction as $E_{T2}$ increases. 
This effect is also present in the theory level curve (where it is seen as
a falling distribution), even for fixed $\alps$, and reflects the 
tendency for the higher 
$E_{T2}$ events to have a smaller $\deta$ due to the presence of the 
$(x_1 x_2 s/p_T^2)^{2 \omega_0}$ factor in the numerator 
(the strong fall off of the parton density functions at large $x$ means that
larger $p_T$ tends to occur without a corresponding increase in mean $x$).
We note that we have used the BFKL formalism to sum the
leading logarithms in the numerator of the gap fraction (supplemented with
parton showering and hadronisation as determined by \herwig) whilst using 
established Monte Carlos to predict the denominator. A strictly leading
logarithmic calculation would also sum the leading logarithms in the
denominator which could give rise to a factor in the denominator of 
$\sim (x_1 x_2 s/p_T^2)^{\omega_0}$. This physics has not yet been
incorporated in \herwig\ or \pythia\ and would have the effect of slowing
down the fall of the $E_T$ spectrum with increasing $E_T$.

\begin{figure}[ht]
\begin{minipage}[t]{0.475\textwidth}
\centerline{\resizebox{8cm}{!}{\rotatebox{0}{\includegraphics{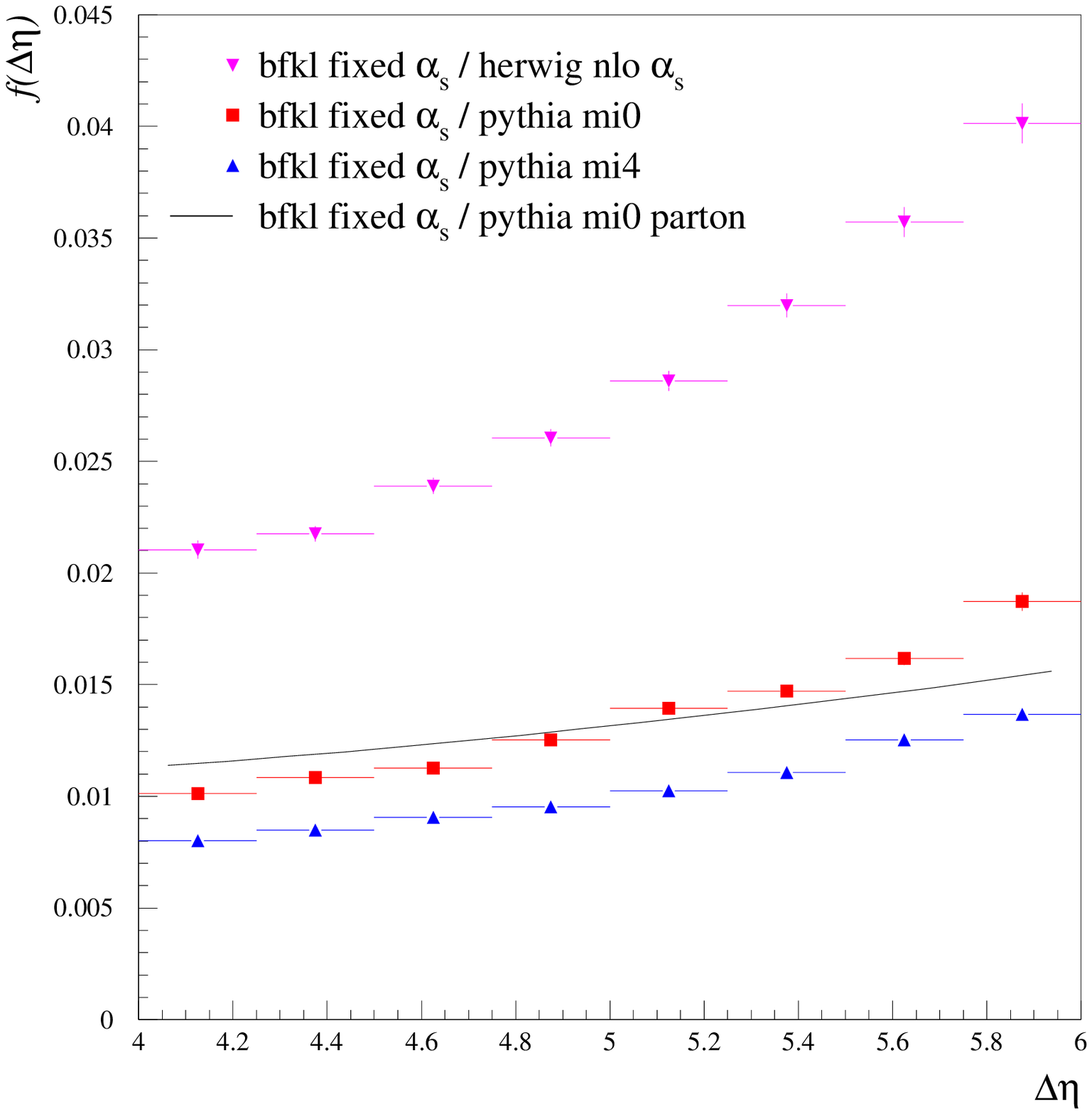}}}}
\caption{Gap fraction as a function of $\Delta \eta$ for 
$15 < E_{T2} < 25$~GeV}
\label{gfetlo}
\end{minipage}\hspace*{\fill}
\begin{minipage}[t]{0.475\textwidth}
\centerline{\resizebox{8cm}{!}{\rotatebox{0}{\includegraphics{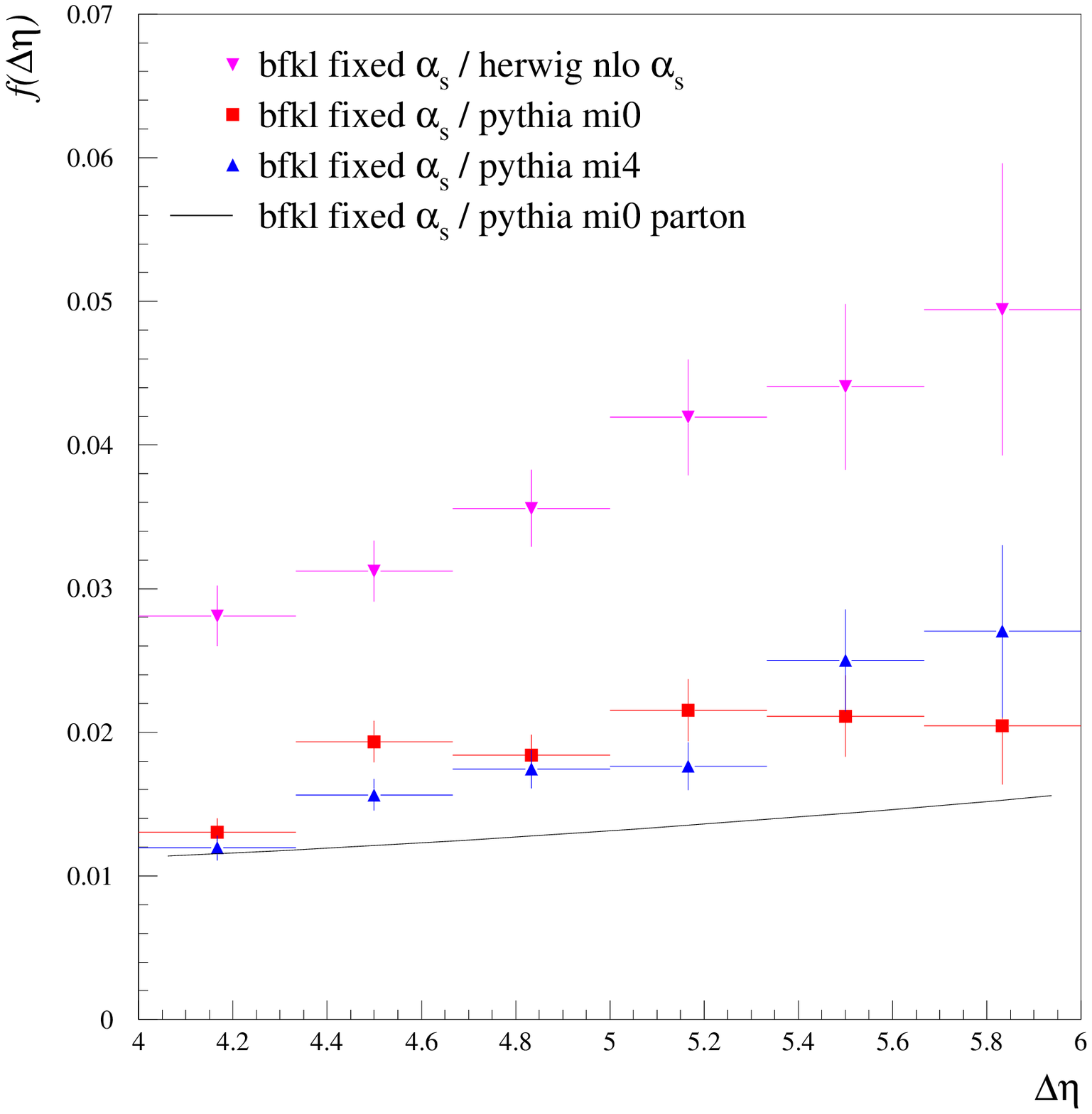}}}}
\caption{Gap fraction as a function of $\Delta \eta$ for $E_{T2} > 30$~GeV}
\label{gfethi}
\end{minipage}
\end{figure}



Figure \ref{gfetlo} shows the gap fraction as a function of $\deta$ for the
low $E_T$ sample and Figure \ref{gfethi} shows it for the high $E_T$ sample.
Notice that the effect of multiple interactions on the denominator of the
gap fraction diminishes as $E_{T2}$ increases. 

Collectively, these gap fraction plots illustrate the uncertainty in
theoretical prediction even in the case of a particular ansatz for the
BFKL behaviour: parton showering and hadronisation corrections are 
significant in computing the gap fraction.

\subsection*{Comparison to data}
Now we come to our comparison with the D\O\ data. In order to make the
comparison, we must first correct our gap events as discussed in Section
\ref{sec:mi}, i.e. we subtract 1~GeV from each jet in the BFKL sample and
generate the non-BFKL sample using \pythia\ without multiple interactions.
 
\begin{figure}[h] 
\centerline{\epsfig{file=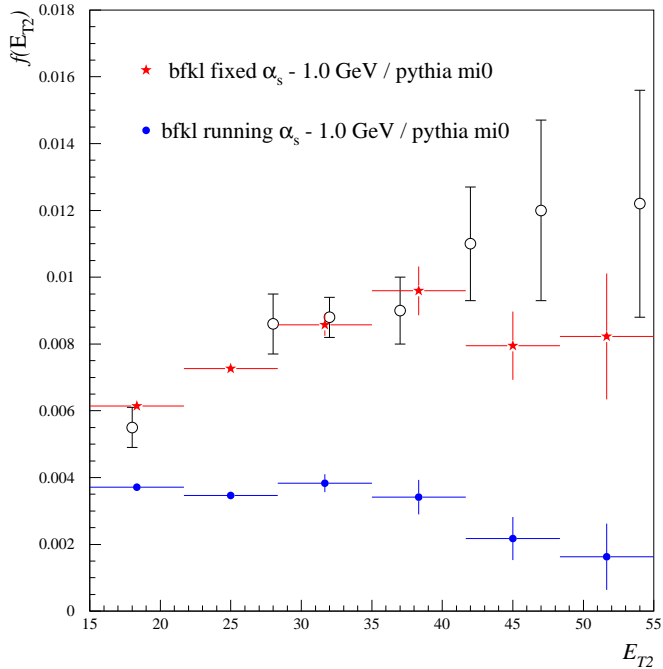,height=10cm}}
\caption{Gap fraction compared to the D\O\ data: $E_{T2}$ spectrum.}
\label{gf:data}
\end{figure}

\begin{figure}[ht]
\begin{minipage}[t]{0.475\textwidth}
\centerline{\resizebox{8cm}{!}{\rotatebox{0}{\includegraphics{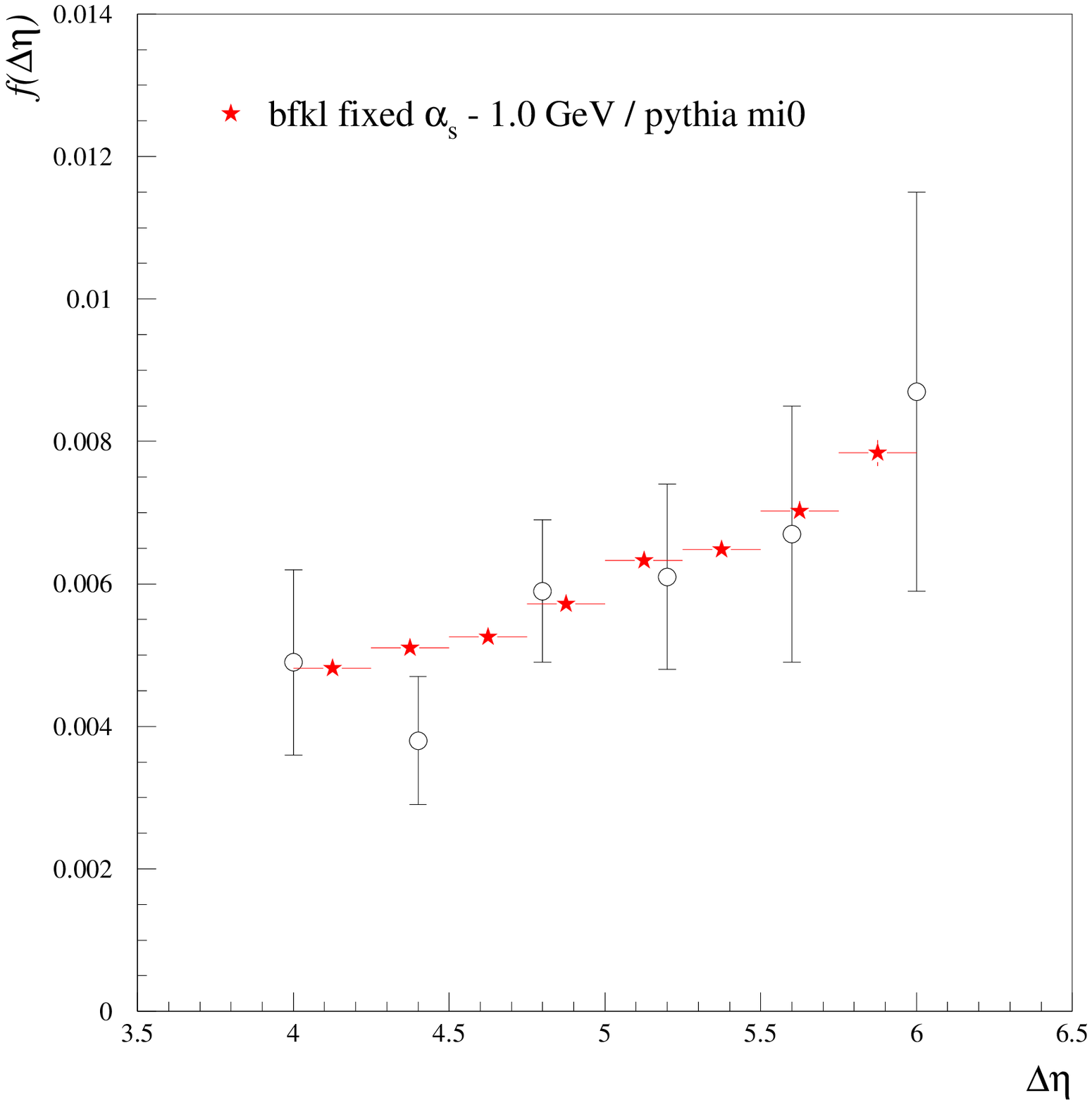}}}}
\caption{Gap fraction compared to the D\O\ data: $\deta$ spectrum at 
low $E_T$.}
\label{gf:loet}
\end{minipage}\hspace*{\fill}
\begin{minipage}[t]{0.475\textwidth}
\centerline{\resizebox{8cm}{!}{\rotatebox{0}{\includegraphics{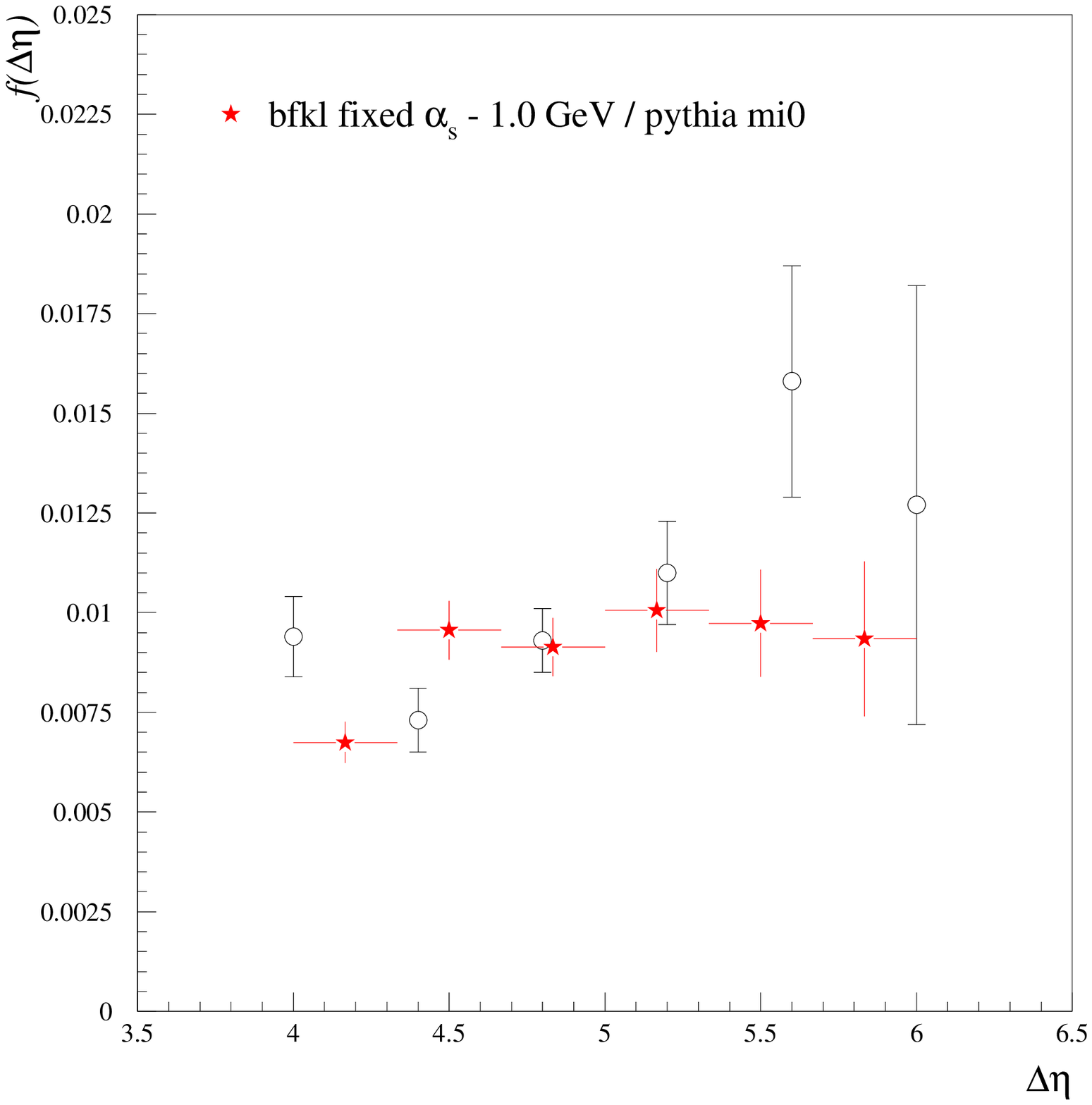}}}}
\caption{Gap fraction compared to the D\O\ data: $\deta$ spectrum at 
high $E_T$.}
\label{gf:hiet}
\end{minipage}
\end{figure}



Figure \ref{gf:data} shows that after fixing the coupling and making the
1~GeV jet correction the open stars agree well with the data. Note that we
have chosen to renormalise our results by a factor 0.6 in those plots where we
compare to the D\O\ data. That this is a reasonable thing to do can be 
appreciated once it is realised that our results have not been fitted to the 
data and that the overall normalisation is acutely sensitive to the magnitude 
of $\alps$. Furthermore, the overall normalisation of the BFKL cross-section 
is uncertain since, within the leading logarithmic approximation, we are free 
to add any small constant to $y$ in (\ref{exact}), i.e. the transformation 
$(s/p_T^2)^{2 \omega} \to (s/(c p_T^2))^{2 \omega}$ where $c$ is some unknown 
constant $\sim 1$ is perfectly admissible. Given these points, we conclude 
that the D\O\ data are in agreement with the leading order BFKL result. 
We note that it is not possible to obtain agreement with the data using 
a strong coupling which runs with the jet $E_T$, even after making the 1~GeV 
subtraction (the solid circles in Figure \ref{gf:data}). 

We could have chosen to use \herwig\ to model the non-BFKL sample. In which
case, the only significant effect, as illustrated in the preceding plots, 
would be a further overall shift in normalisation. 
 
\begin{figure}[ht]
\begin{minipage}[t]{0.475\textwidth}
\centerline{\resizebox{8cm}{!}{\rotatebox{0}{\includegraphics{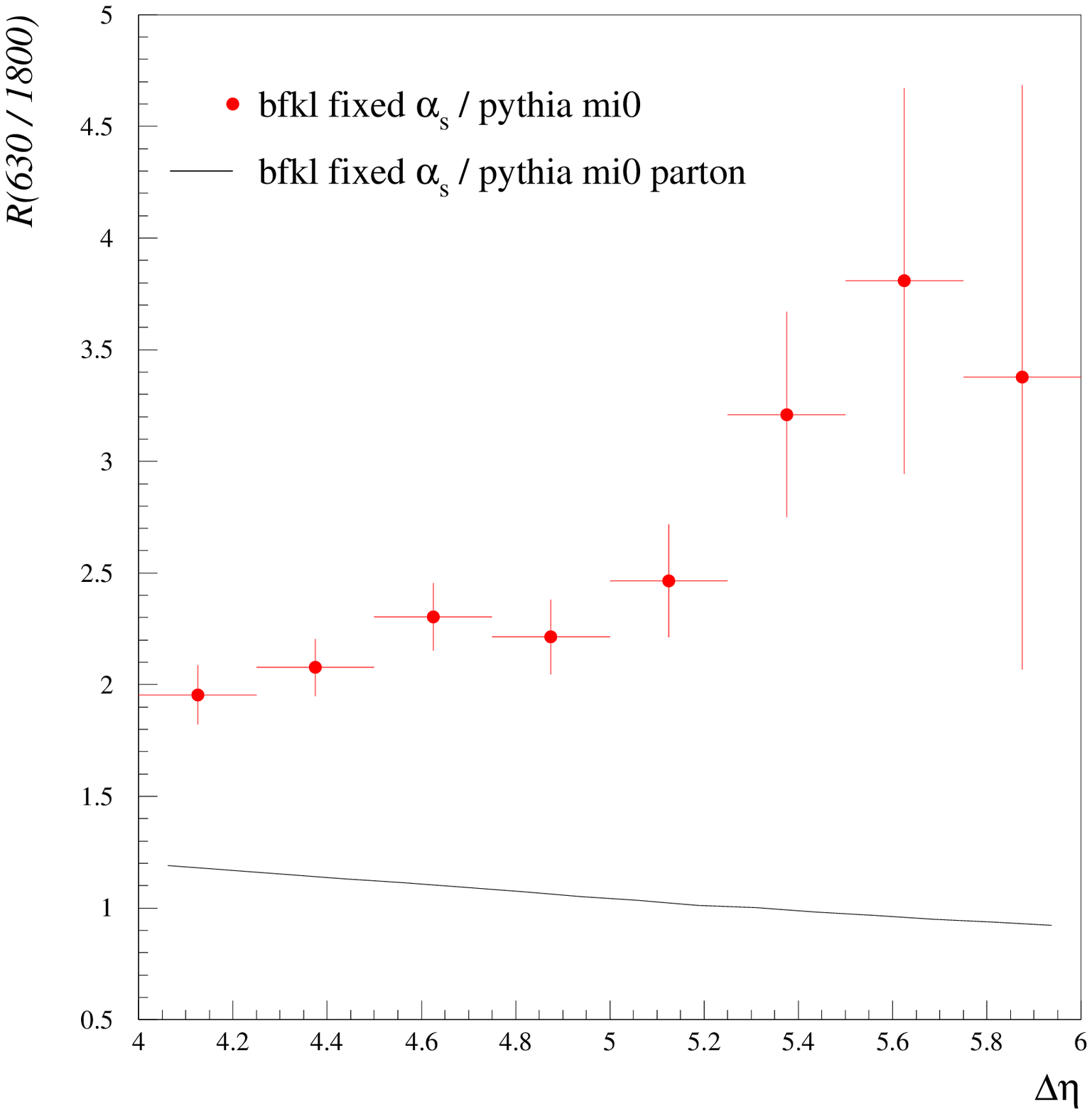}}}}
\caption{Ratio of the gap fraction at 630~GeV to that at 1800~GeV: 
$\deta$ spectrum for $E_{T2} > 15$~GeV.} 
\label{gfrat}
\end{minipage}\hspace*{\fill}
\begin{minipage}[t]{0.475\textwidth}
\centerline{\resizebox{8cm}{!}{\rotatebox{0}{\includegraphics{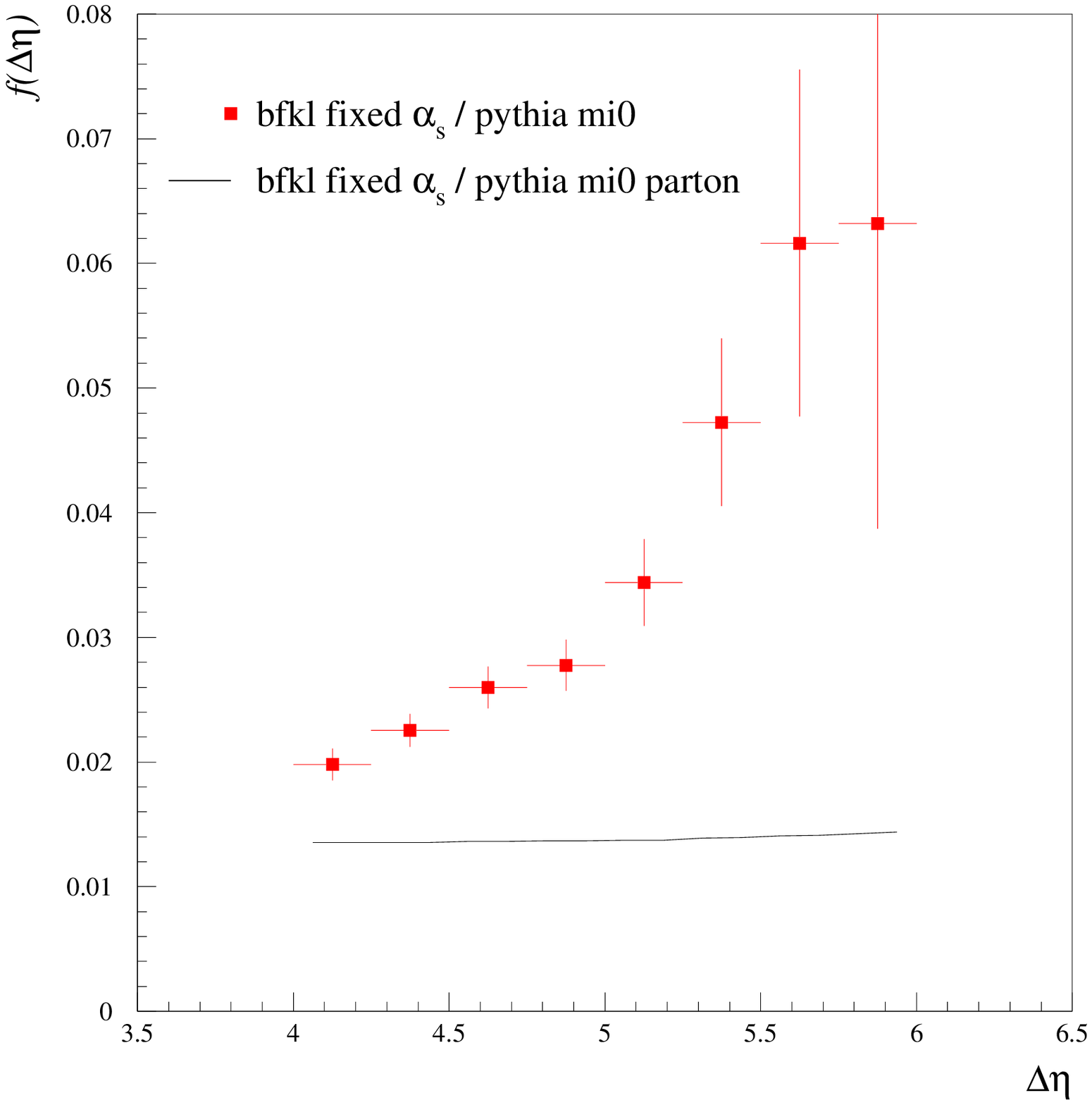}}}}
\caption{Gap fraction at 630~GeV: $\deta$ spectrum for $E_{T2} > 15$~GeV.}
\label{gfrat630}
\end{minipage}
\end{figure}



Figure \ref{gf:loet} shows the $\deta$ distribution at low $E_{T}$ and again
the shape agrees well with the data. A similar story can be seen in 
Figure \ref{gf:hiet} which shows the $\deta$ distribution at high $E_T$. 
The data hint at the possibility that a renormalisation somewhat closer to
unity (compared to the lower $E_T$ case) of the BFKL prediction is
needed in order to bring agreement with the data. However, as we mentioned 
earlier, summing the $\ln 1/x$ terms in the denominator will reduce
the fall of the $E_T$ spectrum as $E_T$ increases and this would in turn
lift the theory predictions in Figure \ref{gf:hiet}. 

In Figure \ref{gfrat} we show our prediction for the D\O\ gap
fraction at 630~GeV compared to that at 1800~GeV. D\O\ has measured the
ratio of the number of gap events at 630~GeV to the number at 1800~GeV and 
finds a value of $3.4 \pm 1.2$ \cite{D01}.\footnote{We note that the D\O\ 
ratio was actually obtained for $E_{T2} > 12$ GeV.}  
This value can be seen to accord with the results shown in
Figure \ref{gfrat}. The data points represent our prediction whilst the solid 
line represents the parton level theory prediction. Clearly the effects of 
parton showering and hadronisation are very significant. 
The rise of the ratio illustrated in Figure \ref{gfrat} at large $\deta$
is understood. The restriction $x<1$ forces the gap and non-gap
cross-sections to fall to zero at some maximum $\deta$, $\deta_{{\rm max}}$.
Now, the colour connection which exists between the jets in the non-gap 
sample drags the jets closer together in rapidity. This has a small effect 
away from $\deta_{{\rm max}}$ (since the $\deta$ spectrum is roughly flat)
however as $\deta \to \deta_{{\rm max}}$ it leads to a more rapid vanishing
of the non-gap cross-section than occurs in the gap cross-section. This
effect, combined with the fact that 
$\deta_{{\rm max}}(630~{\rm GeV}) < \deta_{{\rm max}}(1800~{\rm GeV})$
leads to the rise illustrated in Figure \ref{gfrat}. In Figure \ref{gfrat630}
we show explicitly the gap fraction at 630~GeV and one can see the large
fragmentation corrections which occur as $\deta \to \deta_{{\rm max}}$.
We note that CDF has also measured the ratio as a 
function of the momentum fractions of the final state jets (which are 
approximately equal to the momentum fractions of the incoming 
partons). They find a ratio of $2.4 \pm 0.9$ at fixed momentum fractions (the
value is independent of the momentum fraction) \cite{CDF1}. The fall of
the solid line (representing the tree level prediction) in Figure \ref{gfrat}
arises since the 630~GeV data are at higher parton $x$ than the corresponding
1800~GeV data and are consequently in a region where the gluon density 
falls off more sharply with increasing $x$ than the quark distributions. 
Since the gluon is given more weight in the BFKL sample, this fall compensates
the rise of the gap fraction due to the $\exp(2 \omega_0 y)$ factor leading
to a flatter $\deta$ distribution at 630~GeV than at 1800~GeV.  

\begin{figure}[h] 
\centerline{\epsfig{file=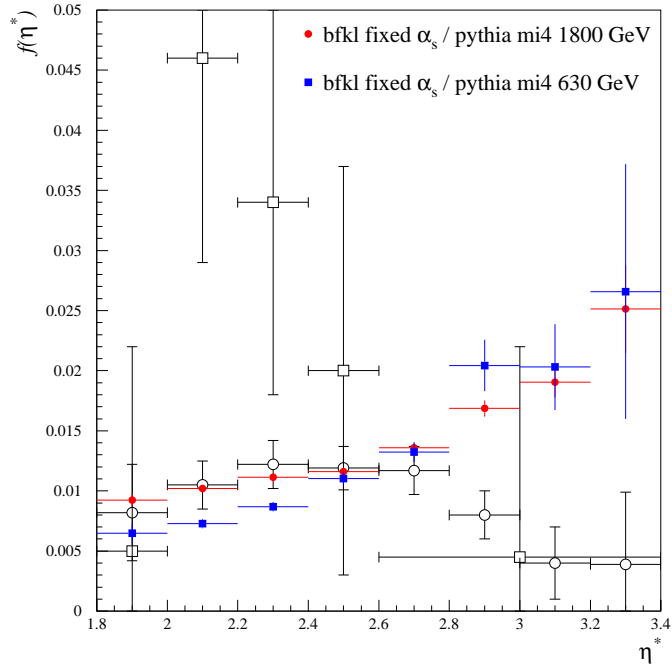,height=10cm}}
\caption{Gap fraction compared to the CDF data at 630 GeV (open squares) and
1800 GeV (open circles).}
\label{gf:cdf}
\end{figure}

Finally, in Figure \ref{gf:cdf} we compare our results with the CDF data
at 630~GeV and 1800~GeV: $\eta^* = \deta/2$. Note that CDF do 
not attempt to correct their jets to include the effect of an underlying
event. In this plot, our theory points are obtained using a renormalisation 
factor of unity (compared to 0.6 in the D\O\ case) which is more in 
line with the normalisation suggested by the higher $E_T$ D\O\ data and, as
we have already discussed, is possibly expected on theoretical grounds
after the inclusion of BFKL effects in the denominator of the gap fraction. 
We then find reasonable agreement with the data except at the larger values of 
$\eta^*$ where we are quite unable to explain a fall in the $\eta^*$ 
distribution. Recall however that D\O\ do not see a fall at large $\deta$.
Further clarification of the situation will require an increase in statistics.

\section{Conclusions}
\label{sec:conc}
We have explicitly demonstrated that the Tevatron data on the gaps-between-jets
process at both 630~GeV and 1800~GeV are in broad agreement with the 
predictions obtained using the leading order BFKL formalism. However, we
are not able to explain the behaviour of the CDF gap fraction at large
$\deta$. Agreement is obtained using the same fixed value of $\alps = 0.17$ 
as was used to explain the recent HERA data on high-$t$ double diffraction 
dissociation.

This agreement should be set in context. The theoretical formalism used does 
suffer from being evaluated only to leading logarithmic accuracy. There is
an essentially unknown overall normalisation (due to the ambiguity in the
choice of scale which defines the leading logarithms, i.e. the parameter $c$
discussed in Section 3) and the procedure for determining the correct
treatment of the running coupling is not defined. To improve upon these
shortcomings requires an understanding of BFKL dynamics at non-zero $t$ which
goes beyond the leading logarithmic approximation.

Crucial in developing a description which is simultaneously consistent with 
the Tevatron data at 630~GeV and at 1800~GeV, and the HERA data at 200~GeV, 
is a realistic treatment of the soft underlying event and its impact on the 
survival of rapidity gaps. Using \pythia\ to model the underlying event we
find an overall gap survival factor which is independent 
of the jet $E_T$ and rapidity but dependent upon the overall centre-of-mass 
energy.

However, the gap fraction is not a very clean observable. We found very 
significant corrections from parton showering and hadronisation to both the 
colour-singlet and non-colour singlet exchange processes which do not cancel 
on taking the ratio and which persist to the largest values of $E_T$. 
Moreover, the influence of the soft underlying event (which we accounted for 
using a multiple interactions model in \pythia), is also very significant at 
the smaller values of $E_T$, i.e. $E_T < 30$~GeV. A good understanding
of the role of the underlying event for $E_T < 30$~GeV is therefore crucial in 
comparing with the data.

\section*{Acknowledgements}
We should like to thank Andrew Brandt, Dino Goulianos, Mark Hayes, Mike 
Seymour and Torbj\"orn Sj\"ostrand for helpful discussions. 
This work was supported by the EU Fourth Framework Programme 
`Training and Mobility of Researchers', Network `Quantum
Chromodynamics and the Deep Structure of Elementary Particles',
contract FMRX-CT98-0194 (DG 12-MIHT). BC would like to thank the UK's
Particle Physics and Astronomy Research Council for support.

\end{document}